\documentclass[aps,prl,twocolumn,showpacs]{revtex4-1}
\usepackage{graphics}
\usepackage{subfigure}
\usepackage{longtable}
\usepackage{graphicx}
\usepackage{pstricks}
\usepackage{amsmath}
\usepackage{dcolumn}
\usepackage{bm}
\usepackage{pdfpages}
\newcommand{\kbr}{$\kappa$-(ET)$_2$Cu[N(CN)$_2$]Br}
\newcommand{\ki}{$\kappa$-(ET)$_2$Cu[N(CN)$_2$]I}
\newcommand{\kpp}{$\kappa^{\prime \prime}$-(ET)$_2$Cu[N(CN)$_2$]Cl}
\newcommand{\kcn}{$\kappa$-(ET)$_2$Cu$_2$(CN)$_3$}
\newcommand{\bi}{$\beta$-(ET)$_2$I$_3$}

\begin{document}

\title{Influence of molecular conformations on the electronic structure of organic charge transfer salts}

\author{Daniel Guterding}
\email{guterding@itp.uni-frankfurt.de}
\affiliation{Institut f\"ur Theoretische Physik, Goethe-Universit\"at Frankfurt, 
Max-von-Laue-Stra{\ss}e 1, 60438 Frankfurt am Main, Germany}

\author{Roser Valent\'i}
\affiliation{Institut f\"ur Theoretische Physik, Goethe-Universit\"at Frankfurt, 
Max-von-Laue-Stra{\ss}e 1, 60438 Frankfurt am Main, Germany}

\author{Harald O. Jeschke}
\affiliation{Institut f\"ur Theoretische Physik, Goethe-Universit\"at Frankfurt, 
Max-von-Laue-Stra{\ss}e 1, 60438 Frankfurt am Main, Germany}

\begin{abstract}
  We report {\it ab-initio} calculations for the electronic structure
  of organic charge transfer salts $\kappa$-(ET)$_2$Cu[N(CN)$_2$]Br,
  $\kappa$-(ET)$_2$Cu[N(CN)$_2$]I, $\kappa^{\prime
    \prime}$-(ET)$_2$Cu[N(CN)$_2$]Cl and
  $\kappa$-(ET)$_2$Cu$_2$(CN)$_3$.
These materials show an ordering of the relative orientation
of terminal ethylene groups in the BEDT-TTF molecules at finite
temperature and our calculations correctly predict the experimentally observed
ground state molecular conformations (eclipsed or staggered).
 Further, it was recently
  demonstrated that the ethylene endgroup relative orientations
  can be used to reversibly tune
  $\kappa$-(ET)$_2$Cu[N(CN)$_2$]Br through a metal-insulator
  transition. Using a tight-binding analysis, we show that the
  molecular conformations of ethylene endgroups are intimately connected
  to the electronic structure and significantly influence hopping
  and Hubbard repulsion parameters. Our results place
 $\kappa$-(ET)$_2$Cu[N(CN)$_2$]Br in eclipsed and staggered configurations
 on opposite sides of the metal-insulator transition.
\end{abstract} 


\maketitle

%Introduction section
Quasi-two dimensional charge transfer salts $\kappa$-(BEDT-TTF)$_2X$, where 
BEDT-TTF stands for bis-ethylenedithio-tetrathiafulvalene, often abbreviated as 
ET, constitute a fascinating family of materials due to their rich phase 
diagrams comprising metallic, superconducting, Mott insulating and spin-liquid 
phases~\cite{Elsinger2000, Shimizu2003, Kurosaki2005, Kagawa2005}.

\begin{figure}
\includegraphics[width=\linewidth]{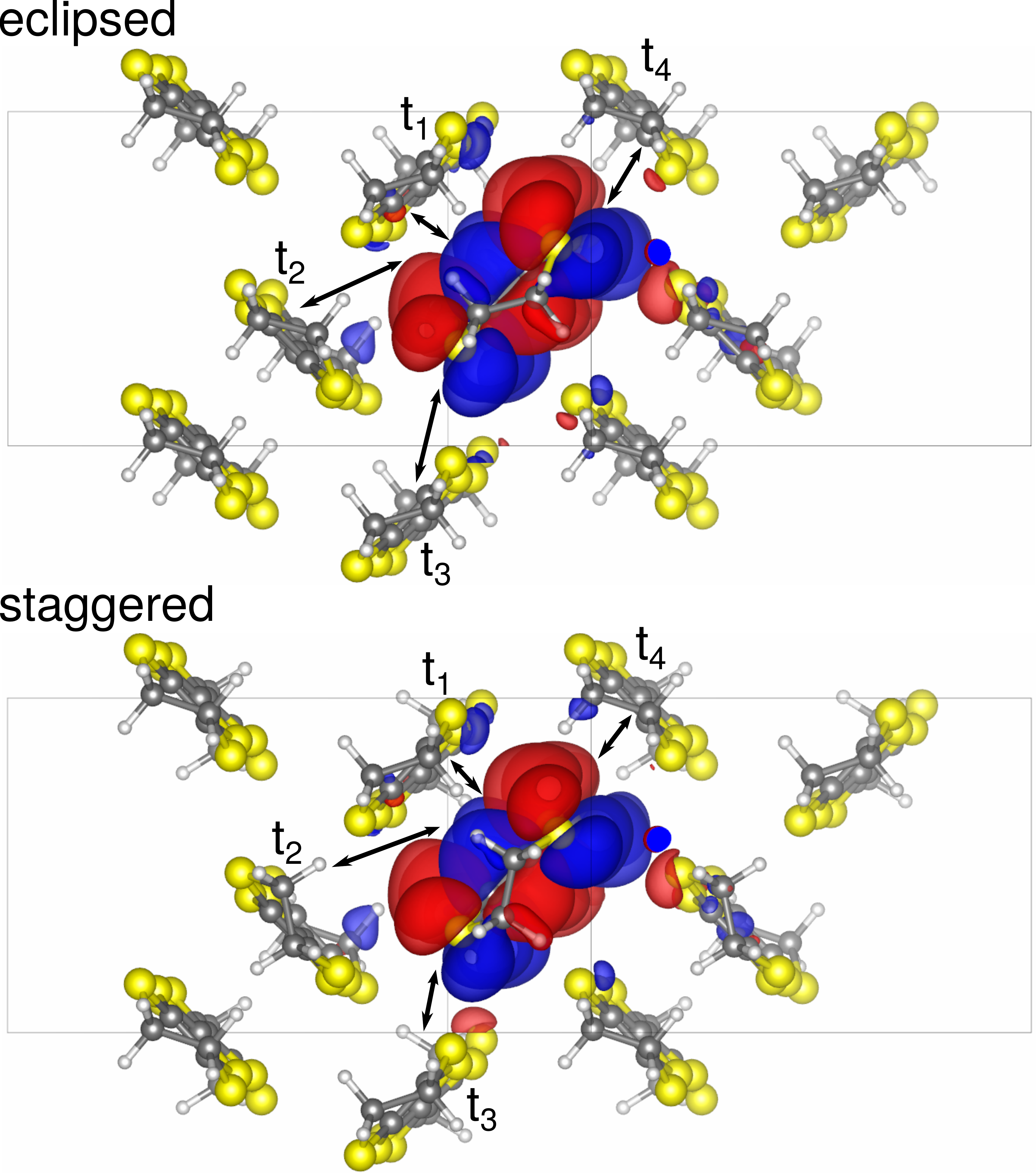}
\caption{(Color online) Molecular Wannier function of an ET molecule
  in staggered and eclipsed configuration for {\kbr}. The arrows
  denote the directions of dominant hopping processes in the
  individual molecule model with ($t_1$,$t_2$,$t_3$,$t_4$).}
\label{fig:wannierfunctions}
\end{figure}

These electronic properties of ET-based materials are very sensitive to disorder.
Irradiation experiments have shown that lattice disorder lowers the
$T_c$ of $\kappa$-(ET)$_2$Cu(SCN)$_2$~\cite{Analytis2006} and causes electron localization
in \kbr~\cite{Sano2010}. Early on it was also realized that ET molecules have intramolecular degrees of freedom, 
namely the configurations of their two ethylene endgroups (see Fig. \ref{fig:wannierfunctions}), which can either be 
aligned parallel (eclipsed, E) or canted (staggered, S)~\cite{Leung1984, 
Leung1985, Schultz1986, Schultz1986PRB, Schirber1986, KET2CuDCAXStructure, 
WatanabePhD, MuellerGlassTransition, Toyota2007, AburtoOrgaz, Mueller2015}. The energetically favorable 
configuration is not universal for different anions $X$ and packing motifs. 
For some materials a glass-like
freezing of the ethylene endgroups upon cooling has 
been observed~\cite{Leung1984, Leung1985,Saito1999,Akutsu2000, Toyota2007, AburtoOrgaz, Mueller2015}.

Especially the first ambient pressure ET-based superconductor \bi ~attracted a 
lot of interest, because its $T_c$ can be enhanced from $1.5~\mathrm{K}$ to 
$8~\mathrm{K}$ by forcing the ET molecules, which are endgroup disordered at 
ambient pressure, to assume staggered configuration through application of shear 
and pressure~\cite{Leung1984, Leung1985, Schultz1986, Schultz1986PRB, 
Schirber1986, Yagubskii1984, Murata1985}. Recently, 
it was shown that ethylene endgroup disorder can be used to 
reversibly tune {\kbr} through a metal to insulator 
transition~\cite{HartmannGlasstransition, HartmannCriticalSlowing, Mueller2015}.

It is believed that materials $\kappa$-(ET)$_2X$ have a common phase
diagram, which is mainly controlled by the value of the on-site
Coulomb repulsion $U$ over the electronic
bandwidth~\cite{KanodaPhaseDiagram}. Changes in physical properties
in the presence of ethylene endgroup disorder have so far been interpreted
as a consequence of lattice disorder, with the exception of recent scanning
tunneling spectroscopy experiments~\cite{Diehl2014}. Surprisingly, the effect of
different ethylene endgroup configurations on the electronic structure, 
and especially the electronic bandwidth,
of $\kappa$-(ET)$_2X$ has only been investigated in a single material
using the extended H\"uckel method~\cite{WatanabePhD}, while
calculations for ET molecules and dimers in vacuum are
available~\cite{ScrivenPowellChem, ScrivenPowell, Mueller2015}. A preceding
study for \kbr~ focused on energetics rather than bandstructure
effects~\cite{AburtoOrgaz}.

Here we examine the electronic structure of endgroup {\it ordered}
crystals in both possible configurations for various members of the
$\kappa$-(ET)$_2$X family of materials using {\it ab-initio} density functional
theory (DFT) calculations. We construct a 3/4-filled
low-energy effective Hamiltonian using a Wannier downfolding scheme
and relate the resulting model parameters to the relative orientation
of terminal ethylene groups. Our results show that ethylene endgroup
configurations of ET molecules influence the electronic bandwidth of
all resulting crystalline materials investigated here. Finally, we
offer a simplified interpretation of our results in terms of the
Hubbard model on the anisotropic triangular lattice and relate our
findings to recent experiments, especially pointing out the possibility that
changes in the electronic structure through ethylene endgroup disorder
and strongly enhanced electron correlation are relevant
in addition to commonly considered lattice disorder.

%Method section
We used {\it ab-initio} density functional theory (DFT) calculations
within an all-electron full-potential local orbital ({\sc
  FPLO})~\cite{FPLOmethod} basis to calculate the electronic
bandstructure. For the exchange-correlation functional we employed the
generalized gradient approximation (GGA)~\cite{PerdewBurkeErnzerhof}.
All calculations were converged on $6\times6\times6$ $k$-point grids.

Tight-binding models were obtained from projective molecular orbital
Wannier functions as implemented in {\sc FPLO}~\cite{FPLOtightbinding}.
We have shown previously that, for the materials of interest here,
this method yields near perfect representations of the low-energy DFT
bandstructure and avoids ambiguities from band fitting
procedures~\cite{DualLayerElectronicStructure}. The resulting
tight-binding Hamiltonian $H = \sum_{ij\sigma} t_{ij}(c^\dagger_{i \sigma} c^{\,}_{j \sigma} + h.c.)$
is 3/4-filled and consists of four bands
corresponding to the four ET sites indexed by $i,j$. Each site is located at
the center of the inner C-C bond of an ET molecule. 
Subsequently, we relate those {\it ab-initio} calculated
parameters with an effective two-band half-filled Hubbard model on the
anisotropic triangular lattice.

Experimental crystal structures including both possible ethylene
endgroup configurations are available~\cite{KET2CuDCAXStructure,
  KppET2CuDCAClStructure, KET2Cu2CN3Structure} for {\ki}, {\kpp} and
{\kcn} ($T = 200~\mathrm{K}$), while for {\kbr} only one endgroup configuration has been
reported in the literature~\cite{KET2CuDCAXStructure}. For the latter
case, we set up the missing ethylene endgroup orientation by hand. For
{\ki} we used the crystal structure determined at $T=127\,\mathrm{K}$
if not denoted otherwise. The structure at $T=295\,\mathrm{K}$ is used
as a consistency check.

\begin{figure}
\includegraphics[width=\linewidth]{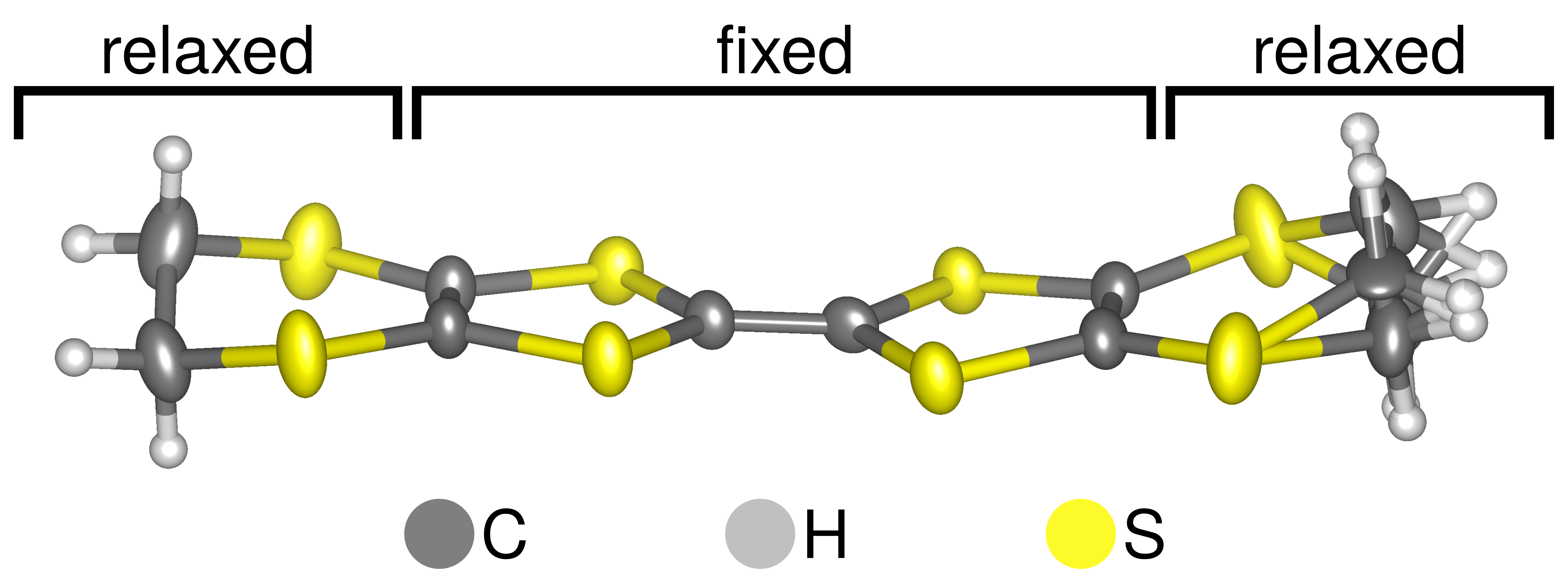}
\caption{(Color online) ET molecule with displacement ellipsoids from
  the experimental structure determination of {\kpp}. The size of the
  ellipsoids scales with the amount of uncertainty regarding the
  respective atomic position. Ethylene endgroup disorder is included
  on the right end of the molecule and given by fractionally occupied
  positions. The bars above the molecule indicate which parts of the
  ET molecules are relaxed in our DFT calculations and which atomic
  positions remain at their experimental values.}
\label{fig:displacement}
\end{figure}

In Figure~\ref{fig:displacement} we show as an example the displacement
ellipsoids resulting from the structure determination of {\kpp}. These
represent the uncertainty of the experimentally determined structure
with respect to the position of every individual atom. Displacements
on the endgroup carbon atoms are largest, but the uncertainty of the
sulphur atom positions next to the ethylene endgroups is also
significant. Therefore, we chose to relax the entire ethylene
endgroups together with the two neighboring sulphur atoms in both
staggered and eclipsed endgroup configurations for all materials
investigated here. For comparison of the relaxed crystal structures
with eclipsed and staggered endgroups see
Fig.~\ref{fig:wannierfunctions}. Other atomic positions and the unit
cell parameters were left untouched. Note that hydrogen positions
belonging to the ethylene groups were not measured in experiment and
had to be inserted manually.

For the structural relaxation we used the projector augmented wave
(PAW) method~\cite{PAWmethod} as implemented in {\sc
  GPAW}~\cite{GPAWmethod}. We optimized the terminal ethylene endgroups
using $2\times2\times2$ $k$-point grids and GGA exchange-correlation
functional until forces were below $10\, \mathrm{meV/\AA}$.

It turns out that both endgroup configurations locally minimize the
total DFT energy, so that stable crystal structures can be obtained
for both staggered and eclipsed orientations from a DFT relaxation.

%Results section
\begin{table}
\caption{Lowest energy configurations of the ethylene endgroups  
  determined from our DFT relaxed structures.  The energy
  difference $\Delta E$ to the high energy configuration is calculated
  per ET molecule from the {\sc FPLO} total energies.}
\begin{ruledtabular}
\begin{tabular}{rcr}
 & configuration & $\Delta E$ in meV\\
\hline
{\kcn} & staggered & 130\\
{\kpp} & eclipsed & 72\\
{\kbr} & eclipsed & 110\\
{\ki} & staggered & 38\\
\end{tabular}
\end{ruledtabular}
\label{tab:endgroupenergies}
\end{table}

It is experimentally known that different charge transfer salts can
either have staggered or eclipsed endgroups as their lowest energy
configuration. To confirm the validity of our relaxed structures, we
first calculate the energy difference $\Delta E$ between staggered and
eclipsed using {\sc FPLO}. The results are displayed in
Table~\ref{tab:endgroupenergies}. Values for the energy differences
calculated with {\sc GPAW} are in good agreement. The energy ordering
of staggered and eclipsed configurations comes out correctly for all
investigated materials with the exception of {\ki} at $295\mathrm{K}$,
where the energy difference is the smallest and the distribution of
endgroup configurations measured experimentally was found to be
51:49~\cite{KET2CuDCAXStructure, KppET2CuDCAClStructure,
  KET2Cu2CN3Structure}. The value for $\Delta E$ we determined is the
energy difference between the two local minima of the energy
corresponding to staggered and eclipsed configurations, which is not to
be confused with the activation energy. The latter denotes the height
of the potential barrier between those minima, which can be
significantly larger than $\Delta E$~\cite{MuellerGlassTransition,
  HartmannGlasstransition}. Instead, our DFT calculated values
constitute a lower bound for the activation energy: {\ki} with the
smallest energy difference is known to be completely endgroup
disordered at room temperature~\cite{KET2CuDCAXStructure} and {\kpp}
contains about 20\% disorder at room
temperature~\cite{KppET2CuDCAClStructure}, while in the other two materials
the amount of endgroup disorder is in a range of few percent. 

\begin{figure*}
\includegraphics[width=\linewidth]{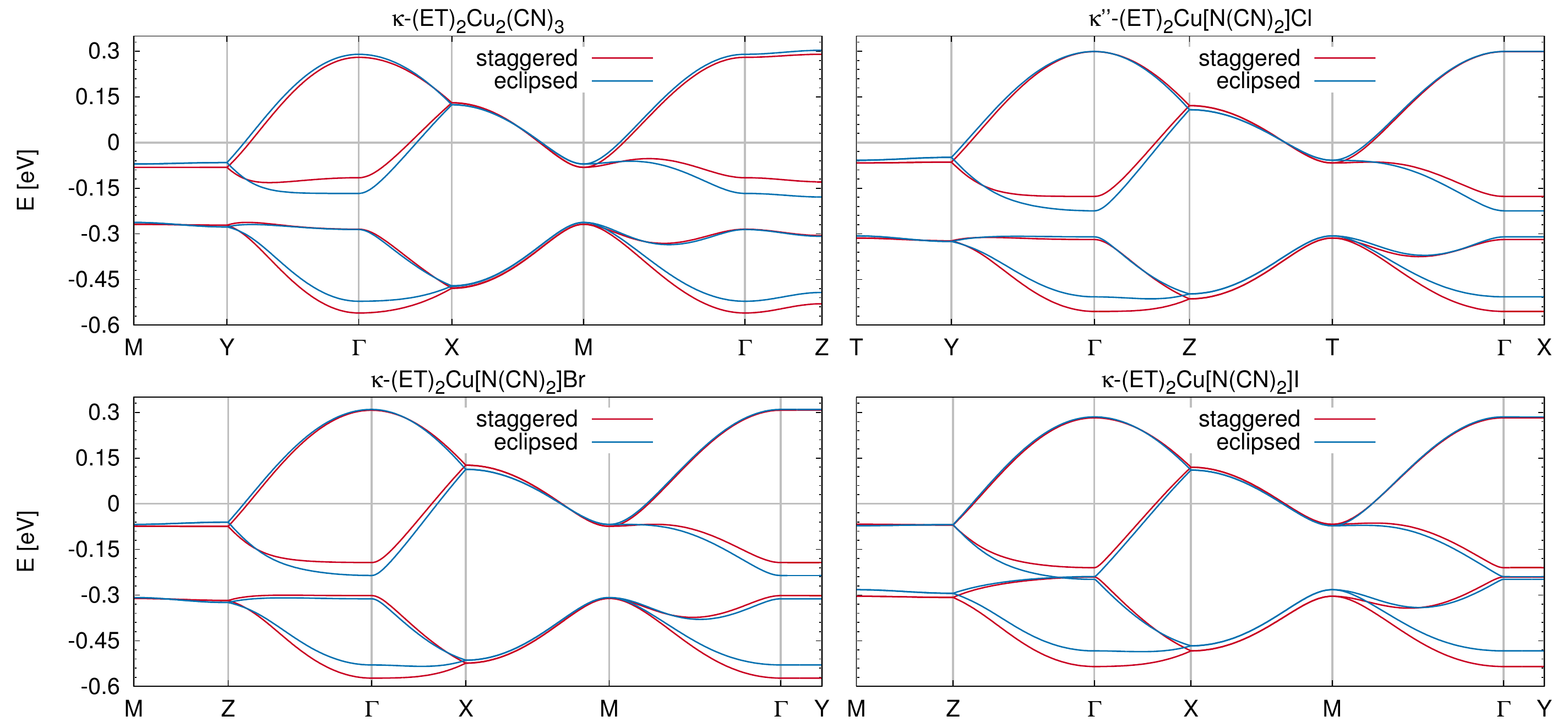}
\caption{(Color online) Electronic bandstructure of all investigated
  materials for staggered and eclipsed ethylene endgroup
  configurations. Staggered ethylene endgroups produce a larger
  overall bandwidth than eclipsed ones, but reduce the width of the
  two bands closest to the Fermi level.}
\label{fig:bandstructure}
\end{figure*}

The electronic bandstructures obtained from the molecular Wannier
function analysis of the DFT results are shown in
Fig.~\ref{fig:bandstructure} for both eclipsed and staggered ethylene
endgroup configurations. In all bandstructures shown, the difference
between staggered and eclipsed configurations lies in the electronic
bandwidth. Going from eclipsed to staggered, the overall bandwidth
increases, while the width of the two bands closest to the Fermi level
decreases.

We analyse these bandstructures using a minimal
model~\cite{KomatsuBandFilling} for a $\kappa$-packed layer of
individual ET molecules (Fig.~\ref{fig:wannierfunctions}) using the
four largest parameters ($t_1$,$t_2$,$t_3$,$t_4$) also commonly
denoted as ($b_1$,$p$,$b_2$,$q$). Note that our tight binding
Hamiltonians also include small longer range hoppings. The four
largest hopping parameters are given in
Tab.~\ref{tab:parametersmolecule}. Parameters $t_1$ ($b_1$) and $t_3$
($b_2$) decrease from staggered to eclipsed configurations, while
$t_4$ ($q$) increases and $t_2$ ($p$) remains about constant.

\begin{table*}
\caption{Values of the molecule model parameters
  ($t_1$,$t_2$,$t_3$,$t_4$) in meV, also commonly
  denoted as ($b_1$,$p$,$b_2$,$q$). Dimer model parameters are given
  as ratios $t^\prime/t$ and $U/t$ calculated from
  ($t_1$,$t_2$,$t_3$,$t_4$) using formulas described in the text. The second 
  column states the experimental ground state of the respective material
  (I=insulator, M=metal, SC=superconductor). The x in the fourth column
  marks the low energy configuration of the ethylene endgroups.}
\begin{ruledtabular}
\begin{tabular}{rrrrrrrr|rr}
 &  &  &  & $t_1$ & $t_2$ & $t_3$ & $t_4$ & $t^\prime/t$ & $U/t$\\
\hline
{\kcn} & I & eclipsed &  & 167 & 84.9 & 70.4 & 30.3 & 0.61 & 5.8 \\
& & staggered & x & 176 & 78.0 & 81.4 & 18.7 & 0.84 & 7.3 \\
\hline
{\kpp} & M  & eclipsed & x & 174 & 97.3 & 50.5 & 35.9 & 0.38 & 5.2\\
& & staggered & & 188 & 93.4 & 64.0 & 26.6 &  0.53 & 6.3 \\
\hline
{\kbr} & SC & eclipsed & x & 178 & 99.0 & 59.5 & 35.8 & 0.44 & 5.3 \\
& & staggered &  & 187 & 97.1 & 70.2 & 24.9 & 0.58 & 6.1 \\
\hline
{\ki}, $127~\mathrm{K}$ & M  & eclipsed & & 152 & 101 & 47.2 & 29.2 & 0.36 & 4.7  \\
& & staggered & x & 170 & 99.0 & 52.4 & 18.9 & 0.44 & 5.8 \\
\hline
{\ki}, $295~\mathrm{K}$ & M  & eclipsed & & 153 & 92.0 & 49.9 & 31.5 & 0.40 & 4.9 \\
& & staggered & & 164 & 92.0 & 54.8 & 22.2 & 0.48 & 5.8  \\
\end{tabular}
\end{ruledtabular}
\label{tab:parametersmolecule}
\end{table*}

The molecular Wannier functions for an ET molecule in eclipsed and
staggered configuration are shown in
Fig.~\ref{fig:wannierfunctions}. Although the molecular Wannier
function hardly resides on the terminal ethylene groups, overlaps with
neighboring ET molecules are influenced by the configuration of the
endgroups through the direction of their bonds with the neighboring
sulphur atoms.

Especially the hopping $t_3$ is strongly enhanced, because the tails
of the Wannier function on the neighboring ET molecule are
enlarged. The Wannier functions remain largely unaltered in the
direction that corresponds to $t_2$. Therefore, this parameter largely
remains constant. In the direction of $t_4$ the tail on the
neighboring ET molecule is enhanced in the staggered configuration,
but the Wannier function on the central molecule turns away from this
tail because of the altered suplhur-ethylene bond
direction. Consequently, $t_4$ is reduced. The relative changes in
$t_1$ are rather small, which is consistent with our analysis of the
Wannier functions.

\begin{figure}
\includegraphics[width=\linewidth]{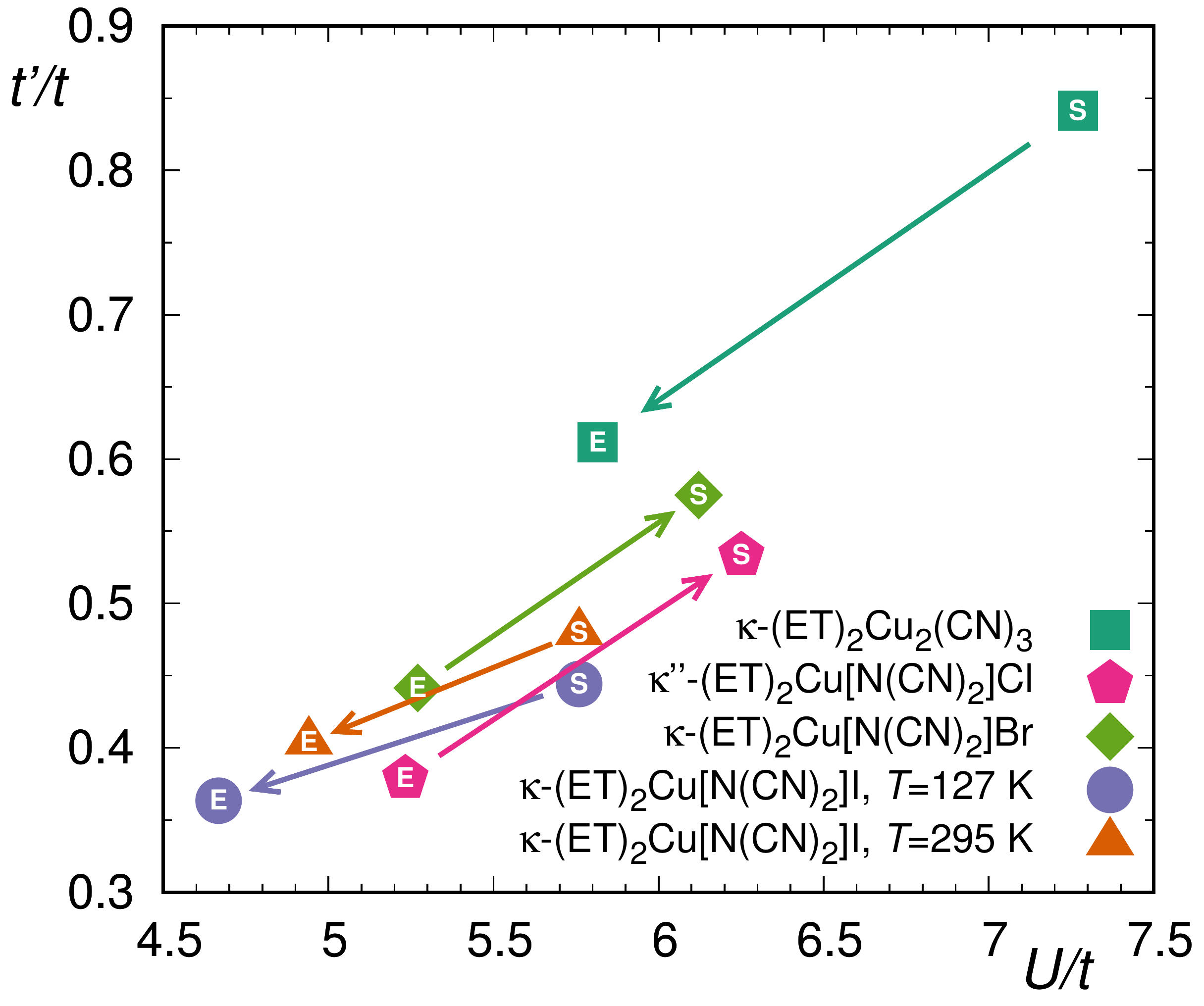}
\caption{(Color online) Parameter ratios $t^\prime / t$ and $U/t$ in
  the effective dimer model. The direction of the arrows indicates the
  direction of change in the model parameters going from the low to
  the high energy configuration of the ethylene endgroups. Eclipsed (E)
  endgroup configurations are located on the left side (small $U/t$)
  of the plot, while staggered (S) endgroup configurations are located on
  the right side (large $U/t$).}
\label{fig:parameters_dimer}
\end{figure}

The effect of the altered hopping integrals and its connection to the
metal-insulator transition in {\kbr} can also be understood in terms
of the simplified two-band dimer model. We calculate the parameters
$t$, $t'$ and $U$ of the corresponding anisotropic triangular lattice
Hubbard from the usual geometric formulas $t = (|t_2|+|t_4|)/2$, $t^\prime
= |t_3| / 2$ and use the rough estimate $U \approx 2 |t_1|$ for the
intradimer Hubbard repulsion~\cite{KomatsuBandFilling, DimerModel}. 

Recently, there has been some criticism of this method of obtaining a $t$, $t'$ 
and $U$ dimer model, because the estimate for $U$ relies on the fact that the 
unscreened intermolecular Coulomb repulsion $V_1$ within an (ET)$_2$ dimer 
vanishes. In DFT parametrizations of this model $V_1$ was however found to be 
non-negligible~\cite{Imamura1999,ScrivenPowellChem,ScrivenPowell}.

Calculations
based on the constrained random phase approximation additionally
showed sizeable screening effects~\cite{NakamuraImada} in the
dimer approximated model not accounted for in our method. In
iron-based materials it is however a well-known problem that
different, even very sophisticated, approximations made in the
determination process can lead to wildly different parameter
sets~\cite{AnisimovDMFT, Aichhorn09, KutepovGW}.

Recent Wannier function analysis anyway points to the fact that a
3/4-filled Hubbard model of individual ET-molecules is more
appropriate to understand fine details of $\kappa$-ET
materials~\cite{KoretsuneHotta, DualLayerElectronicStructure}.
Unfortunately, there are only few examples of many-body calculations
based on such multi-band Hamiltonias~\cite{KurokiRC,Ferber2014}.
Alternatively, a half-filled extended Hubbard model including
non-local Coulomb repulsion and exchange interactions has been 
proposed~\cite{Shinaoka2012}.

We however
emphasize that so far dimer approximated Hamiltonians and subsequent
many-body calculations showed remarkable success in explaining some of
the qualitative properties of ET based
materials~\cite{TremblayDMFT,Clay2008} and should be sufficient for the
case we investigate here. Note that our estimates agree especially well
with more elaborate calculations including screening effects~\cite{NakamuraImada, Kandpal}.
Newly discovered effects like
multiferroic behavior~\cite{Multiferroic} or the still unsettled nature
of superconductivity~\cite{KurokiRC, WosnitzaReview} however certainly call for
more realistic approaches.

Fig.~\ref{fig:parameters_dimer} shows the result of our dimer model
estimate. The change from eclipsed to staggered ethylene group
configuration universally increases both the frustration $t' / t$ and
the relative strength of the Hubbard repulsion $U/t$, i.e. $U$ over
the bandwidth.

Comparison of our findings with cellular dynamical mean-field
theory~\cite{TremblayDMFT} and exact diagonalization results for the
anisotropic triangular lattice Hubbard model~\cite{Clay2008} explains
why {\kbr} can be tuned into a Mott insulating state by activating~\cite{HartmannGlasstransition, HartmannCriticalSlowing} the
energetically less favorable endgroup configuration: First, the
material in its lowest energy configuration is already close to a Mott
insulating phase. Second, the lowest energy configuration is the
eclipsed one, so $U/t$ can be strongly increased by activating the
staggered configuration. Therefore, the system crosses the phase
transition line and a Mott insulator is realized.

Note that the phase diagram of the anisotropic triangular lattice Hubbard
model is not entirely settled and slightly different results have been obtained
using other numerical methods~\cite{Morita2002, Koretsune2007}.

%Summary section
In conclusion, we demonstrated that DFT reliably reproduces the ground state
ethylene endgroup configuration for various $\kappa$-phase materials.
While previous discussion of endgroup conformations in the literature
considered only lattice disorder a relevant issue,
we have shown that the relative orientation of ethylene endgroups
within ET molecules crucially influences the electronic bandwidth of
$\kappa$-type organic charge transfer salts. Switching an ET molecule
from eclipsed to staggered configuration decreases the electronic
bandwidth and in turn enhances the relative strength of the Hubbard
repulsion, bringing the material closer to a Mott insulating
state. Recent experiments where {\kbr} was reversibly switched from a
metallic to an insulating state by tuning the endgroup configurations
are easily understood in our picture. In $\kappa$-type materials that
are not close to any phase transition, the effects of ethylene
endgroup disorder might not manifest as dramatically as in
{\kbr}. Based on our estimates of model parameters, {\kpp} might
exhibit similar behavior. 

%Acknowledgements
\begin{acknowledgments}
  The authors would like to thank Benedikt Hartmann and Jens M\"uller
  for pointing this interesting problem out to us and acknowledge
  support by the Deutsche Forschungsgemeinschaft through grant SFB/TR
  49. Calculations were performed on the LOEWE-CSC and FUCHS
  supercomputers of the Center for Scientific Computing (CSC) in
  Frankfurt am Main, Germany.
\end{acknowledgments}

%Appendix section
%\appendix

\bibliographystyle{apsrev4-1}

\end{document}